\documentclass[11pt,preprint,showpacs,preprintnumbers,amsmath,amssymb]{revtex4}

\usepackage{graphicx}
\usepackage{dcolumn}
\usepackage{bm}
\usepackage{multirow}

\usepackage{CJK}

\begin{document}
\begin{CJK*}{GB}{gbsn}

\title{Spectra of charmed and bottom baryons with hyperfine interaction}

\author{Zhen-Yang Wang (ÍõÕñÑó) \footnote{e-mail: wangz-y@mail.bnu.edu.cn}}
\affiliation{\scriptsize{College of Nuclear Science and Technology, Beijing Normal University, Beijing 100875, China}}

\author{Ke-Wei Wei (κ¿Æΰ) \footnote{Corresponding author, e-mail: weikw@hotmail.com}}
\affiliation{\scriptsize{College of Physics and Electrical Engineering, Anyang Normal University, Anyang 455000, China}}

\author{Jing-Juan Qi (Æî¾´¾ê) \footnote{e-mail: qijj@mail.bnu.edu.cn}}
\affiliation{\scriptsize{College of Nuclear Science and Technology, Beijing Normal University, Beijing 100875, China}}

\author{Xin-Heng Guo (¹ùкã) \footnote{Corresponding author, e-mail: xhguo@bnu.edu.cn}}
\affiliation{\scriptsize{College of Nuclear Science and Technology, Beijing Normal University, Beijing 100875, China}}

\date{\today}

\begin{abstract}

Up to now, the excited charmed and bottom baryon states have still not been well studied experimentally or theoretically. In this paper, we predict the mass of $\Omega_b^*$, the only $L = 0$ baryon state which has not been observed, to be 6069.2 MeV. The spectra of charmed and bottom baryons with the orbital angular momentum $L = 1$ are studied in two popular constituent quark models, the Goldstone boson exchange (GBE) model and the one gluon exchange (OGE) hyperfine interaction model. Inserting the latest experimental data from the ``Review of Particle Physics", we find that in the GBE model, there exist some multiplets ($\Sigma_{c(b)}$, $\Xi'_{c(b)}$ and $\Omega_{c(b)}$) in which the total spin of the three quarks in their lowest energy states is 3/2, but in the OGE model there is no such phenomenon. This is the most important difference between the GBE and OGE models. These results can be tested in the near future. We suggest more efforts to study the excited charmed and bottom baryons both theoretically and experimentally, not only for the abundance of baryon spectra, but also for determining which hyperfine interaction model best describes nature.
\end{abstract}


\maketitle

\section{Introduction}
\label{sec:intro}
The Goldstone boson exchange (GBE) and one gluon exchange (OGE) hyperfine interaction terms describe quark interactions
 in the constituent quark model and are popular for studying baryon spectra \cite{Isgur:1978xj,Glozman:1995fu,Glozman:1995xy,Isgur:1978wd,Copley:1979wj}. These two different kinds of hyperfine interactions have been used to describe the observed spectra of light baryons and ground state heavy baryons \cite{Isgur:1978xj,Maltman:1980er,Glozman:1995fu}. The GBE model can correctly describe the Roper resonance but the OGE model cannot, as stated in Ref. \cite{Mathur:2003zf}. This is a big difference between the GBE and OGE models in light baryons. With the ongoing development of experiments there should be more heavy baryons observed experimentally in the near future, which will in turn guide theoretical studies in this area. The most important motivation of this paper is to compare the differences between the numerical results for negative parity charmed and bottom baryons (with the orbital angular momentum $L = 1$) in these two models.

Baryonic physics in the charmed and bottom sectors has experienced spectacular progress in recent years due to the experimental activities of the BaBar, CLEO, Belle, CDF, and LHCb Collaborations and theoretical developments. Up to now, most of the charmed and bottom baryon ground states have been observed experimentally, but excited heavy baryon states are still poorly known~\cite{Agashe:2014kda}. Recently, the LHCb Collaboration observed four bottom baryon resonances, i.e. $M_{\Lambda_b^{*0}(5912)} = 5911.97$ MeV and $M_{\Lambda_b^{*0}(5920)} = 5919.77$ MeV, which are interpreted as the orbitally excited states of $\Lambda_b^0$ \cite{Aaij:2012da}, and $M_{\Xi_b^{'-}} = 5935.02$ MeV and $M_{\Xi_b^{*-}} = 5955.33$ MeV, which are expected in this mass region with spin-parity $J^P = (1/2)^+$ and $J^P = (3/2)^+$, respectively \cite{Aaij:2014yka}. There are also some states which have been observed, but their $J^P$ numbers have not been determined experimentally. For example, the charmed baryon $\Sigma_c(2800)$ (Belle 2005 \cite{Mizuk:2004yu}) was first reported in the decay modes $\Lambda_c^+\pi^-$, $\Lambda_c^+\pi^0$ and $\Lambda_c^+\pi^+$, with the mass differences $M(\Sigma_c)-M(\Lambda_c^+)$ measured to be $61^{+18+22}_{-13-13}$ MeV for the neutral state, $62^{+37+52}_{-23-38}$ MeV for the charged state, and $75^{+18+12}_{-13-11}$ MeV for the doubly charged state, and was also observed by the BaBar Collaboration in 2008 \cite{Aubert:2008ax}. However, the $J^P$ numbers of $\Sigma_c(2800)$ have not been determined experimentally. There will be more heavy baryons to be observed experimentally in the near future, which can help to examine which hyperfine interaction model can better describe the baryon spectra.

In our paper, we will study the spectra of charmed and bottom baryon states with  negative parity and obital angular momentum $L = 1$. The constituent quark model is a simple and effective phenomenological model to study mass spectra \cite{Godfrey:1985xj,De Rujula:1975ge}. With the data for the light baryons and ground state charmed and bottom baryons which are labeled with three or four stars in the ``Review of Particle Physics" and hence have been well established experimentally, we can calculate the parameters in the constituent quark model. After that, masses of the orbitally excited $L = 1$ charmed and bottom baryons can be calculated. Many calculations based on the quark flavor group $SU(3)$ are remarkably consistent with experiments. When we add another degree of freedom in flavor space, known as $charm$ or $beauty$, a natural generalisation is to extend the flavor group to $SU(4)$ (which means the flavor space we consider is $u$, $d$, $s$, $c$ or $u$, $d$, $s$, $b$ of $SU(4)$), which is however actually badly broken. So in the calculation we introduce a perturbation term which contains two parts. One is the mass difference between the light quark $(u, d)$ and the heavy quark $(c, b)$, and the other is $\Delta_s$, which is from the mass difference between the light quarks ($u$ and $d$) and the $s$ quark. We will neglect the quark mass difference between the light quarks ($u$ and $d$) in the calculation.

The remainder of this paper is organized as follows. In Section \ref{sec:Theo}, we present the theoretical framework, which includes explicit forms of the employed hyperfine interactions between quarks. Numerical results for the spectra of $L = 1$ charmed and bottom baryons are presented in Section \ref{sec:Num}. Finally, Section \ref{sec:Con} contains a brief conclusion.

\section{Theoretical framework}
\label{sec:Theo}

In the constituent quark model, the non-relativistic Hamiltonian for a three-quark system can be expressed as \cite{Glozman:1995fu,Glozman:1995xy}
\begin{equation}\label{H}
{H}=H_0+H_{hyp}+\sum_{i=1}^3m_i,
\end{equation}
where $m_i$ denotes the constituent mass of the $i$th quark, $H_{hyp}$ is the hyperfine interaction between quarks, which is often treated as a perturbation, and $H_0$ is the Hamiltonian concerning orbital motions of the quarks, which should contain two parts, namely the kinetic term and the confining potential between quarks. Both the orbital Hamiltonian $H_0$ and the hyperfine interaction $H_{hyp}$ for the three-quark system have been discussed intensively  in the literature \cite{Isgur:1978xj,Isgur:1978wd,Glozman:1995fu,De Rujula:1975ge}.

\subsection{$H_0$ and the wave-function for a baryon system}
\label{sec:Ham}

The form of $H_0$ employed for the non-relativistic harmonic oscillator potential in the three-quark system is as follows \cite{Glozman:1995fu,Glozman:1995xy}:

\begin{equation}
{H}_{0}=\sum_{i=1}^3\frac{\vec{p}_{i}^2}{2m_{i}}+\sum_{i<j}^3V_{conf}(\vec{r}_{ij}),
\end{equation}
where $\vec{p}_{i}$ and  $m_i$ denote the momentum and mass of the $i$th quark, respectively. The quantity $\vec{r}_{ij} = \vec{r}_i-\vec{r}_j$ is the relative position of the ($ij$) pair of quarks, and $V_{conf}(\vec{r}_{ij})$ is the confinement potential. The harmonic oscillator, as one of the most commonly used quark confinement potentials, has been successfully applied to the spectroscopy of nonstrange and strange baryon ground states and excitations \cite{Glozman:1995fu,Glozman:1995xy}. So we take $V_{conf}(\vec{r}_{ij})$ to be the harmonic oscillator form as follows:
\begin{equation}
V_{conf}(\vec{r}_{ij})=\frac12m{\omega}^2{\vec{r}_{ij}}^2+V_0,
\end{equation}
where $m$ is the constituent mass of the light quarks ($u$, $d$), $\omega$ is the angular frequency of the oscillator interaction and $V_0$ represents the unharmonic part of $V_{conf}$, which is treated as a constant in this paper.

For charmed or bottom baryons, compared with a system including three light quarks, we can rewrite $H_0$ for a system including one heavy quark ($c$ or $b$) as the following (here we do not consider the mass difference between the light quarks ($u$, $d$) and $s$ quark):
\begin{equation}
{H}_{0}=\sum_{i=1}^3\frac{\vec{p}_{i}^2}{2m}+\frac{1}{2}\sum_{i<j}^3m{\omega}^2{(\vec{r}_i-\vec{r}_j)}^2+3V_0+H_0^{'},
\end{equation}
where $H_0^{'}$ represents the corrections due to the mass difference between the light quark and the heavy quark.

If we neglect the contribution from the perturbation term $H_0^{'}$, the exact eigenvalue of $H_0$ should be
\begin{equation}
E_0=(N+3)\omega+3V_0,
\label{E}
\end{equation}
where $N$ is the quantum number of the excited state. $\omega$ can be determined from the mass difference between the nucleon and $N(1400)$, as pointed out in Ref.  \cite{Glozman:1995fu}.

For the perturbation term $H_0'$ of charmed or bottom baryons which comes from the heavy quark ($c$ or $b$) mass difference with the light quark, as shown in Ref. \cite{Glozman:1995xy}, we take it to be flavour-dependent,
\begin{equation}
{H}_{0}^{'}=-\frac{m_{h}-m}{2m}\sum_{i=1}^3\frac{\vec{p}_{i}^2}{m_{h}}\delta_{ih},
\label{H'}
\end{equation}
where $m$ and $m_h$ represent the constituent masses of the light and heavy quarks, respectively, and the Kronecker symbol $\delta_{ih}$ is a flavor dependent parameter with value 1 for a heavy quark and 0 for a light quark.

In the constituent quark model which is governed by the above non-relativistic Hamiltonian, we introduce the three-quark wave function, which is factorized into $orbital \otimes colour \otimes flavour \otimes spin$. In this paper, the wave function is described by the Young pattern [$f$], where $f$ is a sequence of integers that indicate the number of boxes in the successive rows of the corresponding Young patterns. The pattern [3] represents a completely symmetric state, [21] is the mixed symmetric state, and [111] is the completely antisymmetric one. Due to the Pauli principle, the wave function of a three-quark system must be totally antisymmetric under the exchange of any quark pair, so it can be written as $[111]_{XCFS}$ with the subscripts $X$, $C$, $F$ and $S$ (we use $S$ to represent the total spin of the three quarks in the following) representing $orbital$, $colour$, $flavour$ and $spin$ degrees of freedom, respectively.

For the $L = 0$ baryon state all the quarks are in the orbital ground state with $[3]_X$ configuration, and for the $L = 1$ state two quarks are in the orbital ground state and the other in the $P$ state with the $[21]_X$ configuration. Because of colour confinement, the colour wave function must be $[111]_C$. There are three possible flavour wave functions for a baryon system: $[3]_F$, $[21]_F$ and $[111]_F$ in the Weyl tableaux of the $SU(3)$ group \cite{chen,ma,Close}. The total spins could be $S = 1/2$ and $S = 3/2$, corresponding to $[21]_S$ and $[3]_S$ configuration, respectively. The explicit wave functions based on the orbital-colour-flavour-spin configurations can  easily be derived from the Clebsch-Gordan coefficients.

In the case of $N = 0$, all the three quarks are in their ground state, so the matrix elements of $H_0^{'}$ are the same for all the $N = 0$ configurations  \cite{Glozman:1995xy}:

\begin{equation}
<g.s.|H_{0}^{'}|g.s.>=-\frac{1}{2}\delta,
\label{H_01}
\end{equation}
where $|g.s.>$ represents the ground state and $\delta=(1-m/{m_{h}})\omega$.

The perturbation (\ref{H'}) is flavor dependent and its matrix elements between different $P$ shell multiplets of $\Lambda_c^{+}$ take the following values  \cite{Glozman:1995xy}:
\begin{equation}
\begin{split}
&<\Lambda_c^+|H_0^{'}|\Lambda_c^+>_{[21]_{FS}[111]_F[21]_S}=-\frac23\delta,\\
&<\Lambda_c^+|H_{0}^{'}|\Lambda_c^+>_{[21]_{FS}[21]_F[21]_S}=-\frac23\delta,\\
&<\Lambda_c^+|H_{0}^{'}|\Lambda_c^+>_{[21]_{FS}[21]_F[3]_S}=-\frac7{12}\delta,
\label{H_02}
\end{split}
\end{equation}
where the subscript $[21]_{FS}[111]_F[21]_S$, for example, means the configuration with $[21]_{FS}$ flavour-spin symmetry, $[111]_F$ flavour wave-function and $[21]_S$ spin wave-function, and similarly for the other subscripts.

For the $P$ shell excitations of $\Sigma_c$, the matrix elements of $H_0^{'}$ are \cite{Glozman:1995xy}
\begin{equation}
\begin{split}
&<\Sigma_c^+|H_{0}^{'}|\Sigma_c^+>_{[21]_{FS}[21]_F[21]_S}=-\frac23\delta,\\
&<\Sigma_c^+|H_{0}^{'}|\Sigma_c^+>_{[21]_{FS}[3]_F[21]_S}=-\frac23\delta,\\
&<\Sigma_c^+|H_{0}^{'}|\Sigma_c^+>_{[21]_{FS}[21]_F[3]_S}=-\frac34\delta.
\label{H_03}
\end{split}
\end{equation}

For the $L = 1$ negative parity excitations of $\Xi_c$, corrections arising from $H_0^{'}$ are the same as those of $\Lambda_c$, and for the negative parity excitations of $\Xi_c^{'}$ and $\Omega_c$, the corresponding corrections are equal to those of $\Sigma_c$.

\subsection{Hyperfine interactions between quarks}

To calculate the mass splittings of the degenerate configurations, explicit perturbative hyperfine interactions are needed. Since the GBE \cite{Glozman:1995fu,Glozman:1995xy} and OGE \cite{De Rujula:1975ge,Isgur:1978wd,Isgur:1979be,Capstick:2000qj,Isgur:1977ef} interactions between quark pairs have been discussed intensively before, we just present a very brief review and apply them to charmed and bottom baryons.

For charmed and bottom baryons, the GBE hyperfine Hamiltonian can be written in the following form, as in Ref. \cite{BorkaJovanovic:2010yc,Buisseret:2011aa}:

\begin{equation}
H_{GBE}={\sum_{i<j}}V_M{\lambda_{i}^{a}}{\lambda_{j}^{a}}{\vec{\sigma}_i}\cdot{\vec{\sigma}_j},
\label{H_{GBE}}
\end{equation}
where $\lambda_{i}^{a}$ ($a=1,\cdots,14$) are the $SU(4)$ extension of the $SU(3)$ Gell-Mann matrices in flavour space, $\sigma_i$ are Pauli spin matrices (the subscribe $i$ and $j$ represent the $i$th and $j$th quarks, respectively), and $V_{M}$ is a flavor dependent parameter to describe the strength of the exchange of a meson $M$ ($M$ contains $\pi$, $K$, $\eta$, $D$, $D_s$, $B$ and $B_s$ mesons). Because $\eta_c$ and $J/\psi$ are purely $c\bar{c}$ mesons, we do not need to consider the fifteenth Gell-Mann matrix $\lambda^{15}$ of $SU(4)$. Explicitly, the hyperfine interaction Eq. (\ref{H_{GBE}}) between two quarks has the following form for the GBE interaction in the case of the $SU(4)$ extension:

\begin{equation}
H_{GBE} = -\sum_{i<j}\Bigg\{\sum_{a=1}^{3}V_\pi\lambda_i^a \lambda_j^a+\sum_{a=4}^{7} V_K\lambda_i^a \lambda_j^a + V_\eta\lambda_i^8\lambda_j^8+\sum_{a=9}^{12}V_D\lambda_i^a \lambda_j^a+\sum_{a=13}^{14} V_{D_s}\lambda_i^a \lambda_j^a\Bigg\}\vec \sigma_i \cdot \vec \sigma_j.
\label{H_{GBE1}}
\end{equation}

For the OGE interaction \cite{Isgur:1978wd}, the commonly used hyperfine interaction can be written as:
\begin{equation}
H_{OGE}={\sum_{i,j}}C_{i,j}{\lambda_{i}^{C}}{\lambda_{j}^{C}}{\vec{\sigma}_i}\cdot{\vec{\sigma}_j},
\label{H_{OGE}}
\end{equation}
where the $\lambda_{i}^{C}$ and $\sigma_i$ are Gell-Mann $SU(3)$ matrices in colour space and Pauli spin matrices, respectively, and $C_{i,j}$ are the colormagnetic interaction strengths.

\section{Numerical results}
\label{sec:Num}
In this section, we present the numerical results for the $L = 1$ charmed and bottom baryon spectra using the hyperfine interactions given in the GBE and OGE models. Before that we should fix the parameters in these models. For the constituent quark masses we take the values from Refs. \cite{BorkaJovanovic:2010yc,BorkaJovanovic:2012mz,Karliner:2014gca}, which are determined by fitting the experimental baryon masses, i.e. $m_u=m_d=360$ MeV, $m_s=530$ MeV, $m_c=1700$ MeV, $m_b=5043$ MeV. The angular frequency is determined from the mass difference between the nucleon and $N(1400)$ \cite{Glozman:1995xy,Glozman:1995fu}, $\omega=157.3$ MeV. All other parameters in these two different hyperfine interaction models will be obtained from the ground baryon state splittings, which will be discussed in the following.

\subsection{Fine structure corrections of the light, charmed and bottom ground baryons}

Generally, the fine structure corrections ($\delta$M) contain three parts, the hyperfine interaction, the difference $\Delta_s$ between the constituent masses of the light quarks ($u$ and $d$) and $s$ quark, and the energy shift in Eq. (\ref{H'}) which is caused by the heavy quark mass difference. For $L = 0$ states, the energy shift is $-\frac12\delta$.
For the GBE and OGE models, all these corrections ($\delta$M) are presented in Table 1 and Table 2, where all the masses of baryons are taken from the ``Review of Particle Physics'' \cite{Agashe:2014kda}, except for $\Omega_b^*$,  which has not been observed experimentally. We calculate the mass of $\Omega_b^*$ below.

\begin{table}
Table 1 Fine structure corrections ($\delta$M) to the masses (in MeV) of the light ground state baryons (L=0) from the GBE interaction and OGE
interaction. The experimental values are from Ref. \cite{Agashe:2014kda}.
\begin{ruledtabular}
\begin{tabular}{l *{4}{l}}
$[f]_{C}[f]_{FS}[f]_{F}[f]_{S}$  &State &$\delta$M (GBE)& $\delta$M (OGE) &Exp.Mass\\
\hline
$[111]_{C}[3]_{FS}[21]_F[21]_S$ & $N(939)$  &  $-14V_\pi-\frac12\delta$ & $-8C_{qq}-\frac12\delta$ &938.92 \\

$[111]_{C}[3]_{FS}[3]_F[3]_S$ &   $\Delta(1232)$ &  $-4V_\pi-\frac12\delta$ & $8 C_{qq}-\frac12\delta$ &1232  \\

$[111]_{C}[3]_{FS}[21]_F[21]_S$ & $\Lambda(1116)$ &  $-8V_\pi-6V_K+\Delta_s-\frac12\delta$ & $-8 C_{qq}+\Delta_s-\frac12\delta$ &  1115.68  \\

$[111]_{C}[3]_{FS}[21]_F[21]_S$  &$\Sigma(1193)$ &  $\frac43V_\pi-\frac{38}3V_K+\Delta_s-\frac12\delta$ & $\frac{8}{3}C_{qq}$+$\frac{-32}{3} C_{qs}+\Delta_s-\frac12\delta$ &  1193.15  \\

$[111]_{C}[3]_{FS}[3]_F[3]_S$ &   $\Sigma^*(1385)$ & $\frac43V_\pi-\frac83V_K+\Delta_s-\frac12\delta$ & $\frac{8}{3}C_{qq}$+$\frac{16}{3} C_{qs}+\Delta_s-\frac12\delta$ &1384.57  \\

$[111]_{C}[3]_{FS}[21]_F[21]_S$ & $\Xi(1318)$ &  $-\frac{38}3V_K-\frac43V_\eta^{ss}+2\Delta_s-\frac12\delta$ & $\frac{-32}{3}C_{qs}$+$\frac{8}{3} C_{ss}+2\Delta_s-\frac12\delta $& 1318.29   \\

$[111]_{C}[3]_{FS}[3]_F[3]_S$ &   $\Xi^*(1530)$ & $-\frac83V_K-\frac43V_{ss}+2\Delta_s-\frac12\delta$ & $\frac{16}{3}C_{qs}$+$\frac{8}{3} C_{ss}+2\Delta_s-\frac12\delta$ &  1533.4 \\

$[111]_{C}[3]_{FS}[3]_F[3]_S$ &   $\Omega^{*-}(1672)$ & $-4V_\eta^{ss}+3\Delta_s-\frac12\delta$& $8 C_{ss}+3\Delta_s-\frac12\delta$ &1672.45 \\

\end{tabular}
\end{ruledtabular}
\end{table}

 In the GBE hyperfine interaction, we assume that $V_\eta^{qq}$ in $qq$ pair state and $V_\eta^{qs}$ in $qs$ pair state are equal to the exchange potential from $\pi$ and $K$ as in Ref. \cite{Glozman:1995fu}, respectively. In the OGE hyperfine interaction, $C_{qq}$ is a flavor dependent strength parameter. The parameters $V_\pi$, $V_K$, $C_{qs}$ and $C_{qq}$ can be obtained from the $N(939) - \Delta(1232)$ and $\Sigma(1193)-\Sigma^*(1385)$ mass splittings:

\begin{equation}
\begin{split}
M_{\Delta(1232)}-M_{N(939)}=10V_\pi=16C_{qq},\\
M_{\Sigma^*(1385)}-M_{\Sigma(1193)}=10V_K=16C_{qs}.
\end{split}
\end{equation}
Therefore,
\begin{equation}
\begin{split}
&V_\pi = 29.31 \textrm{MeV},\qquad C_{qq} = 18.31 \textrm{MeV},\\
&V_K = 20.32 \textrm{MeV},\qquad C_{qs} = 11.96\textrm{MeV}.
\end{split}
\end{equation}

\begin{table}
Table 2 Fine structure corrections ($\delta$M) to the masses (in MeV) of the charmed and bottom ground baryons (L=0) from the GBE interaction and the OGE
interaction. The experimental values are from Ref. \cite{Agashe:2014kda}.
\begin{ruledtabular}
\begin{tabular}{l    *{4}{l}}
$[f]_{C}[f]_{FS}[f]_{F}[f]_{S}$     & State            &$\delta$M (GBE)    & $\delta$M (OGE) &Exp.Mass\\
\hline
$[111]_{C}[3]_{FS}[21]_{F}[21]_{S}$ & $\Lambda_c$      &  $-8V_\pi-6V_D-\frac12\delta$ & $-8 C_{qq}-\frac12\delta$ &2286.46 \\

$[111]_{C}[3]_{FS}[21]_{F}[21]_{S}$ & $\Sigma_c$       & $-\frac43V_\pi-10V_D-\frac12\delta$ &$\frac83C_{qq}-\frac{32}3C_{qc}-\frac12\delta$ &2452.90\\

$[111]_{C}[3]_{FS}[3]_{F}[3]_{S}$   & $\Sigma_c^*$     & $-\frac43V_\pi-4V_D-\frac12\delta$ &$\frac83C_{qq}$+$\frac{16}3C_{qc}-\frac12\delta$ &2517.50 \\

$[111]_{C}[3]_{FS}[21]_{F}[21]_{S}$ & $\Xi_c$          & $-8V_K-3V_D-3V_{D_s}+\Delta_s-\frac12\delta$ &$-8C_{qs}+\Delta_s-\frac12\delta$ &2467.80\\

$[111]_{C}[3]_{FS}[21]_{F}[21]_{S}$ & $\Xi_c^{'}$      & $-\frac43V_K-5V_D-5V_{D_s}+\Delta_s-\frac12\delta$ &$\frac83C_{qs}-\frac{16}3C_{qc}-\frac{16}3C_{sc}+\Delta_s-\frac12\delta$&2575.60\\

$[111]_{C}[3]_{FS}[3]_{F}[3]_{S}$   & $\Xi_c^*$        & $-\frac43V_K-2V_D-2V_{D_s}+\Delta_s-\frac12\delta$ &$\frac83C_{qs}+\frac83C_{qc}+\frac83C_{sc}+\Delta_s-\frac12\delta$ &2645.90\\

$[111]_{C}[3]_{FS}[21]_{F}[21]_{S}$ & $\Omega_c$       & $-\frac43V_\eta^{ss}-10V_{D_s}+2\Delta_s-\frac12\delta$ &$\frac83C_{ss}-\frac{32}3C_{sc+\Delta_s}-\frac12\delta$ &2695.20\\

$[111]_{C}[3]_{FS}[3]_{F}[3]_{S}$   & $\Omega_c^*$     & $-\frac43V_\eta^{ss}-4V_{D_s}+2\Delta_s-\frac12\delta$ &$\frac83C_{ss}+\frac{16}3C_{sc}+\Delta_s-\frac12\delta$ &2765.90\\

$[111]_{C}[3]_{FS}[21]_{F}[21]_{S}$ & $\Lambda_b$      & $-8V_\pi-6V_B-\frac12\delta$ & $-8C_{qq}-\frac12\delta$ & 5619.50\\

$[111]_{C}[3]_{FS}[21]_{F}[21]_{S}$ & $\Sigma_b$       & $-\frac43V_\pi-10V_B-\frac12\delta$ & $\frac83C_{qq}-\frac{32}3C_{qb}-\frac12\delta$ &5813.4\\

$[111]_{C}[3]_{FS}[3]_{F}[3]_{S}$   & $\Sigma_b^*$     & $-\frac43V_\pi-4V_B-\frac12\delta$ & $\frac83C_{qq}+\frac{16}3C_{qb}-\frac12\delta$ &5833.6\\

$[111]_{C}[3]_{FS}[21]_{F}[21]_{S}$ & $\Xi_b$          & $-8V_K-3V_B-3V_{B_s}+\Delta_s-\frac12\delta$ & $-8 C_{qs}+\Delta_s-\frac12\delta$&5794.9 \\

$[111]_{C}[3]_{FS}[21]_{F}[21]_{S}$ & $\Xi_b^{'}$      & $-\frac43V_K-5V_B-5V_{B_s}+\Delta_s-\frac12\delta$& $\frac83C_{qs}-\frac{16}3C_{qb}-\frac{16}3C_{sb}+\Delta_s-\frac12\delta$ &5835.02\\

$[111]_{C}[3]_{FS}[3]_{F}[3]_{S}$   & $\Xi_b^*$        & $-\frac43V_K-2V_B-2V_{B_s}+\Delta_s-\frac12\delta$&  $\frac83C_{qs}+\frac83C_{qb}+\frac83C_{sb}+\Delta_s-\frac12\delta$ &5955.33 \\

$[111]_{C}[3]_{FS}[21]_{F}[21]_{S}$ & $\Omega_b$       & $-\frac43V_\eta^{ss}-10V_{B_s}+2\Delta_s-\frac12\delta$& $\frac83C_{ss} -\frac{32}3C_{sb}+\Delta_s-\frac12\delta$ &6048.8\\

$[111]_{C}[3]_{FS}[3]_{F}[3]_{S}$   & $\Omega_b^*$    & $-\frac43V_\eta^{ss}-4V_{B_s}+2\Delta_s-\frac12\delta$& $\frac83C_{ss}+\frac{16}3C_{sb}+\Delta_s-\frac12\delta$& \\

\end{tabular}
\end{ruledtabular}
\end{table}

To determine the parameters $V_D$, $V_{D_s}$, $C_{qc}$ and $C_{sc}$, we consider the $\Sigma_c - \Sigma_c^*$ and $\Omega_c-\Omega_c^*$ mass splittings:
\begin{equation}
\begin{split}
&M_{\Sigma_c^*}-M_{\Sigma_c}=6V_D=16C_{qc},\\
&M_{\Omega_c^*}-M_{\Omega_c}=6V_{D_s}=16C_{sc}.
\end{split}
\end{equation}
Therefore,
\begin{equation}
\begin{split}
&V_D = 10.77 \textrm{MeV},\qquad C_{qc} = 4.04 \textrm{MeV},\\
&V_{D_s} = 11.78 \textrm{MeV},\qquad C_{sc} = 4.42\textrm{MeV}.
\end{split}
\end{equation}

The mass splitting between $\Sigma_b$ and $\Sigma_b^*$ is:
\begin{equation}
M_{\Sigma_b^*}-M_{\Sigma_b}=6V_B=16C_{qb}.
\end{equation}
Therefore,
\begin{equation}
V_B = 3.37 \textrm{MeV},\qquad C_{qb} = 1.26 \textrm{MeV}.
\end{equation}

Then we consider the $\Xi_b^{*} - \Xi_b^{'}$ mass splitting,
\begin{equation}
M_{\Xi_b^{*}}-M_{\Xi_b^{'}}=3V_B+3V_{B_s}=8C_{qb}+8C_{sb}.
\label{V_B}
\end{equation}
Substituting $V_B = 3.37$ MeV, $C_{qb} = 1.26$ MeV, and the masses of $\Xi_b^{*}$ and $\Xi_b^{'}$ into Eq. (\ref{V_B}), we can get $V_{B_s}=3.40$ MeV and $C_{sb}=1.24$ MeV.
In Refs. \cite{Glozman:1995fu,Glozman:1995xy}, it is pointed out that $V_K = (\frac{m_u}{m_s})V_\pi$, $V_{ss} = (\frac{m_u}{m_s})V_K$. Therefore, $V_{ss} = \frac{V_K^2}{V_\pi} = 14.08 $ MeV. All the parameters in the two different hyperfine interaction models are summarized in Table 3.
\begin{table}
Table 3 Parameters (in MeV) of the two hyperfine interaction models.\label{par}
\begin{ruledtabular}
\begin{tabular}{l    *{8}{l}}

GBE&$V_\pi$ &29.31& $V_K$     &20.32& $V_{ss}$ &14.08&           &     \\

  &$V_D$   &10.77& $V_{D_s}$ &11.78& $V_B$    &3.37 & $V_{B_s}$ &3.40 \\
\hline
OGE&$C_{qq}$&18.31& $C_{qs}$  &11.96& $C_{ss}$ &7.82 &           &     \\

  &$C_{qc}$&4.04 & $C_{sc}$  &4.42 & $C_{qb}$ &1.26 & $C_{sb}$  &1.24\\

\end{tabular}
\end{ruledtabular}
\end{table}

For the $L = 0$ baryon state, $\Omega_b^*$ is the only state which has not been observed experimentally. From Table 2, for both GBE and OGE models, we find that
\begin{equation}\label{equal function}
  (M_{\Omega_b^*}-M_{\Omega_b})+(M_{\Sigma_b^*}-M_{\Sigma_b})=2(M_{\Xi_b^*}-M_{\Xi_b}).
\end{equation}
Note that there is no difference for the relations in Eq. (\ref{equal function}) between these two different hyperfine interaction models. We predict $M_{\Omega_b^*} = 6069.2$ MeV from Eq. (\ref{equal function}). We will discuss the $L = 1$ charmed and bottom baryons in the GBE and OGE models in the next subsection.

\subsection{Masses of charmed and bottom baryon states with L = 1}
\label{sec:charm}

With all the fixed hyperfine interactions parameters and the configurations of the negative parity charmed and bottom baryon systems with $L = 1$ outlined in Table 4, the masses of the charmed and bottom baryon states can be calculated. There are three steps to obtain numerical results in our models. First, one has to calculate the fine structure corrections of charmed and bottom baryon configurations with $L = 1$ from these two different kinds of hyperfine interactions. These can be obtained by calculating the matrix elements of the hyperfine interations in Eq. (\ref{H_{GBE}}) and Eq. (\ref{H_{OGE}}), $\Delta_s$ and the energy shift in Eq. (\ref{H'}). Second, one should calculate the mass of the configurations from the mass splittings between charmed and bottom baryon states with $L = 0$. Finally, by diagonalization of the matrices, we can get the masses of the baryon states.

\begin{table}
Table 4 Flavor-spin configurations for the charmed and bottom baryon systems with $L=1$.
\begin{ruledtabular}
\begin{tabular}{*{3}{l} }
                                  &Configuration      &   Multiplet \\
\hline
  $|\Lambda_{c(b)}\rangle_1$    &$[21]_X[111]_c[21]_{FS}[111]_{F}[21]_{S}$    &  $\frac12^-,\frac32^-,\Lambda_{c(b)}$                     \\

  $|\Lambda_{c(b)}\rangle_2$    &$[21]_X[111]_c[21]_{FS}[21]_{F}[21]_{S}$     &  $\frac12^-,\frac32^-,\Lambda_{c(b)}$                     \\

  $|\Lambda_{c(b)}\rangle_3$    &$[21]_X[111]_c[21]_{FS}[21]_{F}[3]_{S}$      &  $\frac12^-,\frac32^-,\frac52^-,\Lambda_{c(b)}$           \\

  $|\Sigma_{c(b)}\rangle_1$     &$[21]_X[111]_c[21]_{FS}[21]_{F}[21]_{S}$     &  $\frac12^-,\frac32^-,\Sigma_{c(b)}$                     \\

  $|\Sigma_{c(b)}\rangle_2$     &$[21]_X[111]_c[21]_{FS}[3]_{F}[21]_{S}$      &  $\frac12^-,\frac32^-,\Sigma_{c(b)}$                     \\

  $|\Sigma_{c(b)}\rangle_3$     &$[21]_X[111]_c[21]_{FS}[21]_{F}[3]_{S}$      &  $\frac12^-,\frac32^-,\frac52^-,\Sigma_{c(b)}$            \\

  $|\Xi_{c(b)}\rangle_1$        &$[21]_X[111]_c[21]_{FS}[111]_{F}[21]_{S}$    &  $\frac12^-,\frac32^-,\Xi_{c(b)}$                         \\

  $|\Xi_{c(b)}\rangle_2$        &$[21]_X[111]_c[21]_{FS}[21]_{F}[21]_{S}$     &  $\frac12^-,\frac32^-,\Xi_{c(b)}$                         \\

  $|\Xi_{c(b)}\rangle_3$        &$[21]_X[111]_c[21]_{FS}[21]_{F}[3]_{S}$      &  $\frac12^-,\frac32^-,\frac52^-,\Xi_{c(b)}$               \\

  $|\Xi_{c(b)}^{'}\rangle_1$    &$[21]_X[111]_c[21]_{FS}[21]_{F}[21]_{S}$     &  $\frac12^-,\frac32^-,\Xi_{c(b)}^{'}$                     \\

  $|\Xi_{c(b)}^{'}\rangle_2$    &$[21]_X[111]_c[21]_{FS}[3]_{F}[21]_{S}$      &  $\frac12^-,\frac32^-,\Xi_{c(b)}^{'}$                     \\

  $|\Xi_{c(b)}^{'}\rangle_3$    &$[21]_X[111]_c[21]_{FS}[21]_{F}[3]_{S}$      &  $\frac12^-,\frac32^-,\frac52^-,\Xi_{c(b)}^{'}$           \\

  $|\Omega_{c(b)}\rangle_1$     &$[21]_X[111]_c[21]_{FS}[21]_{F}[21]_{S}$     &  $\frac12^-,\frac32^-,\Omega_{c(b)}$                      \\

  $|\Omega_{c(b)}\rangle_2$     &$[21]_X[111]_c[21]_{FS}[3]_{F}[21]_{S}$      &  $\frac12^-,\frac32^-,\Omega_{c(b)}$                      \\

  $|\Omega_{c(b)}\rangle_3$     &$[21]_X[111]_c[21]_{FS}[21]_{F}[3]_{S}$      &  $\frac12^-,\frac32^-,\frac52^-,\Omega_{c(b)}$            \\

\end{tabular}
\end{ruledtabular}
\end{table}

For the $\Lambda_c^+$ multiplet, $\Lambda_c(2595)^+$ and $\Lambda_c(2625)^+$, with $J^P = \frac12^-$ and $J^P = \frac32^-$, respectively, $M_{\Lambda_c(2595)^+}$ = 2592.25 MeV, $M_{\Lambda_c(2625)^+}$ = 2628.11 MeV from the latest ``Review of Particle Physics'' \cite{Agashe:2014kda}. Then we can easily get the mass splitting $M_{\Lambda_c(2625)^+} - M_{\Lambda_c(2595)^+} = 35.86$ MeV. Similarly, for the states $\Xi_c(2790)^+(J^P = \frac12^-)$ and $\Xi_c(2815)^+(J^P = \frac32^-)$, the mass splitting is 27.5 MeV. There are also two orbitally excited singly bottom baryons measured experimentally: $\Lambda_b(5912)$ and $\Lambda_b(5920)$ with $J^P = \frac12^-$ and $J^P = \frac32^-$, respectively, and $M_{\Lambda_b(5912)}$ = 5912.11 MeV, $M_{\Lambda_b(5920)}$ = 5919.81 MeV. The mass difference of these two states is small ($<$ 8 MeV). The spin-orbital interaction has a smaller influence on the mass of the charmed and bottom baryons with the increase of constituent quark masses. As Capstick and Isgur pointed out in Ref. \cite{Capstick:1986bm}, the spin-orbit terms are quite small. So we can safely neglect the impact of spin-orbit coupling on our calculation. Since we have neglected spin-orbital effects, $S$ becomes a good quantum number.

Because $\Lambda_c$ and other states have two configurations with the same total spin $S = 1/2$ as listed in Table 4, we need to consider the mixing of these two configurations. However, for the total spin $S=1/2$ and $S=3/2$ states the mixing is zero because $[21]_S$ is orthogonal to $[3]_S$. Then after explicit derivation, the matrices of fine structure corrections in these two models are:
{\scriptsize{
\begin{equation}
\begin{split}\label{GBE hyperfine}
&{\mathcal{H}_{\Lambda_c}^{GBE}=\left(
\begin{matrix}
-\frac83V_{\pi}-4V_{D}-\frac23\delta    & -\frac89V_{\pi}+\frac23V_{D}            & 0\\
-\frac89V_{\pi}+\frac23V_{D}            & -\frac83V_{\pi}+2V_{D}-\frac 23\delta   & 0\\
          0                             & 0                                       & \frac83V_\pi-2V_D-\frac7{12}\delta
\end{matrix}
\right)=\left(
\begin{matrix}
-139.3    & -18.9     & 0\\
-18.9     &-203.9      & 0\\
  0        & 0          & -110.5
\end{matrix}
\right)},\\
&{\mathcal{H}_{\Sigma_c}^{GBE}=\left(
\begin{matrix}
 \frac43V_{\pi}-2V_{D}-\frac23\delta   &  -\frac89V_{\pi}+\frac43V_{D}          & 0\\
 -\frac89V_{\pi}+\frac43V_{D}           & \frac43V_{\pi}+4V_{D}-\frac 23\delta  & 0\\
           0                            & 0                                      & -\frac43V_\pi+2V_D-\frac34\delta
\end{matrix}
\right)=\left(
\begin{matrix}
-65.1     & -11.7     & 0\\
-11.7     & -0.5      & 0\\
  0        & 0          & -110.5
\end{matrix}
\right)},\\
&{\mathcal{H}_{\Xi_c}^{GBE}=\left(
\begin{matrix}
-\frac83V_K-2V_{D}-2V_{D_s}+\Delta_s-\frac23\delta   &  -\frac49V_K+\frac13V_{D}+\frac13V_{D_s}            & 0\\
          -\frac49V_K+\frac13V_{D}+\frac13V_{D_s}   & -\frac83V_K+V_{D}+V_{D_s}+\Delta_s-\frac23\delta  & 0\\
                     0                              & 0                                                   & \frac83V_K-V_D-V_{D_s}+\Delta_s-\frac7{12}\delta
\end{matrix}
\right)=\left(
\begin{matrix}
-11.9    & -1.5   & 0\\
-1.5     & 55.7   & 0\\
  0       & 0       & 129.3
\end{matrix}
\right)},\\
&{\mathcal{H}_{\Xi_c'}^{GBE}=\left(
\begin{matrix}
\frac43V_K-V_D-V_{D_s}+\Delta_s-\frac23\delta  &  -\frac89V_K+\frac23V_{D}+\frac23V_{D_s}           & 0\\
-\frac89V_K+\frac23V_{D}+\frac23V_{D_s}        & \frac43V_K+2V_D+2V_{D_s}+\Delta_s-{\frac23}\delta  & 0\\
                0                              & 0                                                  & -\frac43V_K+V_D+V_{D_s}+\Delta_s-\frac34\delta
\end{matrix}
\right)=\left(
\begin{matrix}
91.9     & -3.0    & 0\\
-3.0     & 159.5   & 0\\
  0       & 0        & 72.5
\end{matrix}
\right)},\\
&{\mathcal{H}_{\Omega_c}^{GBE}=\left(
\begin{matrix}
\frac43V_\eta^{ss}-2V_{D_s}+2\Delta_s-\frac23\delta  &  -\frac49V_\eta^{ss}+\frac43V_{D_s}                         & 0\\
-\frac49V_\eta^{ss}+\frac43V_{D_s}                         &  \frac43V_\eta^{ss}+4V_{D_s}+2\Delta_s-\frac23\delta  & 0\\
                0                                    & 0                                                     & -\frac43V_\eta^{ss}+2V_{D_s}+2\Delta_s-\frac34\delta
\end{matrix}
\right)=\left(
\begin{matrix}
253.8    & 9.0     & 0\\
9.0      & 324.5   & 0\\
  0       & 0        & 250.5
\end{matrix}
\right)},
\end{split}
\end{equation}}}
where, for example, the matrix $(\mathcal{H}_{\Lambda_c}^{GBE})_{ij}$ is the element of the matrix of $_i\langle \Lambda_c|H_{GBE} + H_0' + n\Delta_s|\Lambda\rangle_j$, and $n$ is the number of $s$ quarks in the baryon state.

{\scriptsize{\begin{equation}\begin{split}\label{OGE hyperfine}
&{\mathcal{H}_{\Lambda_c}^{OGE}=\left(
\begin{matrix}
-\frac83(C_{qq}+2C_{qc})-\frac23\delta  &  \frac83(C_{qq}-C_{qc})                 &   0\\
 \frac83(C_{qq}-C_{qc})                 & -\frac83(C_{qq}+2C_{qc})-\frac23\delta  &   0\\
                0                       &            0                            &   \frac83(C_{qq}+2C_{qc})-\frac7{12}\delta
\end{matrix}
\right)=\left(
\begin{matrix}
-153.0    & 38.1     & 0\\
38.1      & -153.0   & 0\\
  0        & 0         & -1.9
\end{matrix}
\right)},\\
&{\mathcal{H}_{\Sigma_c}^{OGE}=\left(
\begin{matrix}
 -\frac83(2C_{qc}+C_{qq})-\frac23\delta  &  \frac{16}3(C_{qq}-C_{qc})           &  0\\
  \frac{16}3(C_{qq}-C_{qc})              & -\frac83(2C_{qc}+C_{qq})-\frac23\delta &  0\\
                0                        &            0                         &  \frac83(2C_{qc}+C_{qq})-\frac34\delta
\end{matrix}
\right)=\left(
\begin{matrix}
-153.0    & 76.1     & 0\\
76.11      & -153.0   & 0\\
  0        & 0         & -22.6
\end{matrix}
\right)},\\
&{\mathcal{H}_{\Xi_c}^{OGE}=\left(
\begin{matrix}
-\frac83(C_{us}+C_{uc}+C_{sc})+\Delta_s-\frac23\delta  &  \frac43(2C_{qs}-C_{qc}-C_{sc})                         &  0\\
 \frac43(2C_{qs}-C_{qc}-C_{sc})                        & -\frac83(C_{us}+C_{uc}+C_{sc})+\Delta_s-\frac23\delta   &  0\\
                0                                      &            0                                            &  \frac83(C_{us}+C_{uc}+C_{sc})+\Delta_s-\frac7{12}\delta
\end{matrix}
\right)=\left(
\begin{matrix}
32.9     & 20.6     & 0\\
20.6     & 32.9     & 0\\
  0        & 0         & 152.1
\end{matrix}
\right)},\\
&{\mathcal{H}_{\Xi_c'}^{OGE}=\left(
\begin{matrix}
-\frac83(C_{us}+C_{uc}+C_{sc})+\Delta_s-\frac23\delta  &  -\frac83(2C_{qs}-C_{qc}-C_{sc})                        &  0  \\
-\frac83(2C_{qs}-C_{qc}-C_{sc})                        & -\frac83(C_{us}+C_{uc}+C_{sc})+\Delta_s-\frac23\delta   &  0  \\
                0                                      &            0                                            &  \frac83(C_{us}+C_{uc}+C_{sc})+\Delta_s-\frac34\delta
\end{matrix}
\right)=\left(
\begin{matrix}
32.9       & -41.2    & 0\\
-41.2      & 32.9     & 0\\
  0         & 0         & 131.5
\end{matrix}
\right)},\\
&{\mathcal{H}_{\Omega_c}^{OGE}=\left(
\begin{matrix}
-\frac83(2C_{sc}+C_{ss})+2\Delta_s-\frac23\delta  &  \frac{16}3(C_{ss}-C_{sc})                        &  0  \\
 \frac{16}3(C_{ss}-C_{sc})                        & -\frac83(2C_{sc}+C_{ss})+2\Delta_s-\frac23\delta  &  0  \\
                0                                 &            0                                      &  \frac83(2C_{sc}+C_{ss})+2\Delta_s-\frac34\delta
\end{matrix}
\right)=\left(
\begin{matrix}
210.2       & 23.5    & 0\\
23.5      & 210.2     & 0\\
  0         & 0         & 294.1
\end{matrix}
\right)}.
\end{split}
\end{equation}}}

From Eqs. (\ref{H}) to (\ref{H'}) and the corresponding calculation above, we can find that the eigenvalues for the configurations are:
\begin{equation}\label{eigenvalues}
  E = \sum_{i=1}^3m_i + (N + 3)\omega + 3V_0 + \langle H_{GBE(OGE)}\rangle + \langle H'\rangle + n \Delta_s,
\end{equation}
where $m_i$ denotes the constituent mass of the $i$th quark, and $N$ and $n$ represent the quantum number of the excited state and the number of $s$ quarks in the baryon state, respectively. Then, the expressions for the mass splittings between the $L = 1$ $\Lambda_c$ and $L = 0$ $\Lambda_c$ states are:

\begin{equation}\label{Lambda_c1}
  M_{|\Lambda_{c}\rangle_1}^{L=1} - M_{\Lambda_{c}}^{L=0}=\left\{
   \begin{aligned}
   &\frac{16}3V_\pi+2V_D-\frac 16\delta+\omega,  \\
   &\frac{16}3C_{qq}-\frac{16}3C_{qc}-\frac 16\delta+\omega, \\
   \end{aligned}
   \right.
  \end{equation}

\begin{equation}\label{Lambda_c2}
  M_{|\Lambda_{c}\rangle_2}^{L=1} - M_{\Lambda_{c}}^{L=0}=\left\{
   \begin{aligned}
   &\frac{16}3V_\pi+2V_D-\frac 16\delta+\omega,  \\
   &\frac{16}3C_{qq}-\frac{16}3C_{qc}-\frac 16\delta+\omega, \\
   \end{aligned}
   \right.
  \end{equation}

\begin{equation}\label{Lambda_c3}
  M_{|\Lambda_{c}\rangle_3}^{L=1} - M_{\Lambda_{c}}^{L=0}=\left\{
   \begin{aligned}
   &\frac{32}3V_\pi+4V_D-\frac 1{12}\delta+\omega,  \\
   &\frac{32}3C_{qq}+\frac{16}3C_{qc}-\frac 1{12}\delta+\omega, \\
   \end{aligned}
   \right.
  \end{equation}
where the first (second) line in each equation is the result in the GBE (OGE) model.

Inserting the parameters listed in Table 3 and the values for $M_{\Lambda_{c}}^{L=0}$ into the above expressions, one can easily get the masses $M_{|\Lambda_{c}\rangle_{1,2,3}}^{L=1}$. Similarly, we can also obtain the masses of $\Sigma_c$, $\Xi_c$, $\Xi_c^{'}$, $\Omega_c$ states with $L = 1$ by considering the mass splittings between them and their corresponding ground states, which are listed in Table 2. These masses just represent the energies of the configurations and are listed in the diagonal terms of matrices (\ref{GBE results}) and (\ref{OGE results}), but are not the real charmed baryons' physical masses, which will be calculated later. All the input masses of corresponding ground states are taken from the latest ``Review of Particle Physics" \cite{Agashe:2014kda}.

Then we can get matrices $\langle H\rangle$ (H represents the non-relativistic Hamiltonian for a three-quark system) for every multiplet configuration, and the numerical values are listed in the following matrices:
{\scriptsize{
\begin{equation}
\begin{split}\label{GBE results}
&{\mathcal{E}_{\Lambda_c}^{GBE}=\left(
\begin{matrix}
2600.9    & -18.9     & 0\\
-18.9     &2665.6     & 0\\
  0        & 0          & 2789.2
\end{matrix}
\right)},
\quad{\mathcal{E}_{\Sigma_c}^{GBE}=\left(
\begin{matrix}
2753.8    & -11.7     & 0\\
-11.7     & 2818.5    & 0\\
  0        & 0          & 2708.4
\end{matrix}
\right)},\\
&{\mathcal{E}_{\Xi_c}^{GBE}=\left(
\begin{matrix}
2735.4   & -1.5     & 0\\
-1.5     & 2803.0   & 0\\
  0       & 0         & 2893.9
\end{matrix}
\right)},\quad{\mathcal{E}_{\Xi_c'}^{GBE}=\left(
\begin{matrix}
2856.6     & -3.0    & 0\\
-3.0     & 2924.3   & 0\\
  0       & 0        & 2831.5
\end{matrix}
\right)},\quad{\mathcal{E}_{\Omega_c}^{GBE}=\left(
\begin{matrix}
2966.1    & 9.0     & 0\\
9.0      & 3036.8   & 0\\
  0       & 0        & 2962.8
\end{matrix}
\right)},
\end{split}
\end{equation}}}

{\scriptsize{\begin{equation}\begin{split}\label{OGE results}
&{\mathcal{E}_{\Lambda_c}^{OGE}=\left(
\begin{matrix}
2499.2    & 38.1     & 0\\
38.1      & 2499.2   & 0\\
  0        & 0         & 2650.3
\end{matrix}
\right)},\quad{\mathcal{E}_{\Sigma_c}^{OGE}=\left(
\begin{matrix}
2513.4    & 76.1     & 0\\
76.1      & 2513.4   & 0\\
  0        & 0         & 2643.8
\end{matrix}
\right)},\\
&{\mathcal{E}_{\Xi_c}^{OGE}=\left(
\begin{matrix}
2645.7      & 20.6     & 0\\
20.6        & 2645.7   & 0\\
  0          & 0         & 2896.7
\end{matrix}
\right)},\quad{\mathcal{E}_{\Xi_c'}^{OGE}=\left(
\begin{matrix}
2671.0     & -41.2    & 0\\
-41.2      & 2671.0   & 0\\
  0         & 0         & 2839.8
\end{matrix}
\right)},\quad{\mathcal{E}_{\Omega_c}^{OGE}=\left(
\begin{matrix}
2808.4    & 23.5     & 0\\
23.5      & 2808.4   & 0\\
  0        & 0         & 2962.8
\end{matrix}
\right)}.
\end{split}
\end{equation}}}

In the GBE model, from matrix (\ref{GBE results}), one can see the lowest energy states of three multiplets ($\Sigma_c$, $\Xi_c'$ and $\Omega_c$) have spin 3/2. According to our analysis, this is because the contributions from fine structure corrections to $\Sigma_c$, $\Xi_c'$ and $\Omega_c$ states with spin 3/2 are smaller than those states with spin 1/2 in the GBE model from the matrix (\ref{GBE hyperfine}). However, in the OGE model there is no such phenomenon. This is the most important difference between the GBE and OGE models.

By diagonalizing matrices (\ref{GBE results}) and (\ref{OGE results}), we can get the energies for the physical charmed baryon states as shown in Table 5, in which we also show these states as the linear combinations of the configurations given in Table 4, with the corresponding coefficients for the combinations listed in Table 5. From Table 5 we can also see the spin of the states with the lowest energy is also $S = 3/2$ for $\Sigma_c$, $\Xi_c'$ and $\Omega_c$ states only in the GBE model, but not in the OGE model.

\begin{table}
Table 5 Energies and coefficients for mixing between the configurations with $S = 1/2$ and $S = 3/2$ for the charmed baryon states with $L = 1$ in the GBE and OGE models.\label{coe1}
\begin{ruledtabular}
\begin{tabular}{l   *{13}{l}}
GBE&   &$|\Lambda_{c}\rangle_1$&$|\Lambda_{c}\rangle_2$& $|\Lambda_{c}\rangle_3$& &$|\Sigma_{c}\rangle_1$&$|\Sigma_{c}\rangle_2$&$|\Sigma_{c}\rangle_3$ &  &$|\Xi_{c}\rangle_1$&$|\Xi_{c}\rangle_2$&$|\Xi_{c}\rangle_3$\\
\hline
   &2595.8&0.965 &0.261&0   & 2748.7 &  0.965 & -0.261&0    & 2735.3 & 0.999 & 0.022&0     \\

   &2670.7&-0.261&0.965&0   & 2823.6 & 0.261 & 0.965&0   & 2803.0 & -0.022 & 0.999&0     \\

   &2789.2&0    &0     & 1  & 2708.4 &0       & 0     &1   & 2893.9 &0      &0      &1    \\
\hline
   &   &$|\Xi_{c}^{'}\rangle_1$&$|\Xi_{c}^{'}\rangle_2$&$|\Xi_{c}^{'}\rangle_3$& &$|\Omega_{c}\rangle_1$&$|\Omega_{c}\rangle_2$&$|\Omega_{c}\rangle_3$  &     & \\
\hline
   &2856.5& 0.999 & 0.045 &0    & 2964.9  & 0.992 & 0.125 &0  & &   \\

   &2924.4& -0.045 & 0.999&0    & 3037.9  & 0.125 & 0.992 &0  & &  \\

   &2831.5&0       &0      &1    & 2962.8  &0      &0      &1       \\
\hline\hline
OGE&   &$|\Lambda_{c}\rangle_1$&$|\Lambda_{c}\rangle_2$& $|\Lambda_{c}\rangle_3$& &$|\Sigma_{c}\rangle_1$&$|\Sigma_{c}\rangle_2$&$|\Sigma_{c}\rangle_3$ &  &$|\Xi_{c}\rangle_1$&$|\Xi_{c}\rangle_2$&$|\Xi_{c}\rangle_3$\\
\hline
   &2461.2 &0.707 &-0.707&0   & 2437.3 & 0.707 & -0.707&0    & 2619.5 & 0.707 & -0.707&0    \\

   &2537.2 &0.707 & 0.707&0   & 2589.5 & 0.707 & 0.707&0     & 2660.7 & 0.707 & 0.707&0     \\

   &2650.3 &0     &0     & 1  & 2643.8 &0       & 0     &1   & 2896.7 &0      &0      &1    \\
\hline
   &   &$|\Xi_{c}^{'}\rangle_1$&$|\Xi_{c}^{'}\rangle_2$&$|\Xi_{c}^{'}\rangle_3$& &$|\Omega_{c}\rangle_1$&$|\Omega_{c}\rangle_2$&$|\Omega_{c}\rangle_3$   &    & \\
\hline
   &2629.8& 0.707 & -0.707&0    & 2784.9 & 0.707 & -0.707&0    &   &  &  \\

   &2712.2& 0.707 & 0.707&0    & 2831.8 & 0.707 & 0.707&0    &   &    &  \\

   &2839.8&0       &0      &1    & 2962.8  &0      &0      &1       \\
  \end{tabular}
\end{ruledtabular}
\end{table}

 As shown in Table 5, in the GBE model, the mixing of the configurations in the GBE model is much weaker than that in the OGE model. For instance, for the $\Lambda_c$ states, the mixing coefficient between the configurations $|\Lambda_{c(b)}\rangle_1$ and $|\Lambda_{c(b)}\rangle_2$ is about 0.26, but in the OGE model it is about 0.71. According to our calculation results in matrix (\ref{OGE results}), the diagonal matrix elements and nondiagonal matrix elements have the same results when the spin is 1/2 in the OGE model. So the mixing coefficients between the configurations with spin 1/2 should be the same and the mixing is stronger than in the GBE model,as shown in Table 5. The absolute values of the nondiagonal matrix elements in the OGE model are larger than those in the GBE model.

For the bottom baryon states, the fine structure correction matrices are analogous to the expressions in (\ref{GBE hyperfine}) and (\ref{OGE hyperfine}); we just need change the $c$ quark, $D$ and $D_s$ mesons to $b$ quark, $B$ and $B_s$ mesons, respectively. The expressions for the mass splittings between the negative parity bottom baryon states with $L = 1$ and the corresponding states of $L = 0$ are similar to Eqs. (\ref{Lambda_c1}), (\ref{Lambda_c2}) and (\ref{Lambda_c3}). We will not give the explicit expressions for these mass splittings here. The numerical results for the bottom baryon configurations with $L = 1$ are listed in matrices (\ref{M_3}) and (\ref{M_4}), which are obtained in the same way as for the charmed baryons in the GBE and OGE models. We also consider the bottom baryon configuration mixing, and the masses and corresponding coefficients for the mixing are listed in Table 6. There are two differences, however, between bottom baryon states and charmed baryon states. The first is that the larger constituent mass of the $b$ quark leads to the increase of the mass difference correction in Eq. (\ref{H'}), and the second is that the hyperfine interaction contributions from the GBE interaction Eq. (\ref{H_{GBE}}) and the OGE interaction Eq. (\ref{H_{OGE}}) to the cases of the bottom baryon states should be less important than those for the charmed baryon states in these two models. This is because the parameters as listed in Table 3 for bottom baryon states are smaller than for charmed baryon states.

{\scriptsize{
\begin{equation}
\begin{split}
&{\mathcal{E}_{\Lambda_b}^{GBE}=\left(
\begin{matrix}
5915.5    & -18.8     & 0\\
-18.8     &5935.7     & 0\\
  0        & 0          & 6090.7
\end{matrix}
\right)},
\quad{\mathcal{E}_{\Sigma_b}^{GBE}=\left(
\begin{matrix}
6051.5    & -23.8     & 0\\
-23.8     & 6071.7    & 0\\
  0        & 0          & 5974.6
\end{matrix}
\right)},\\
&{\mathcal{E}_{\Xi_b}^{GBE}=\left(
\begin{matrix}
6042.9   & -6.8     & 0\\
-6.8     & 6063.2   & 0\\
  0       & 0         & 6079.1
\end{matrix}
\right)},\quad{\mathcal{E}_{\Xi_b'}^{GBE}=\left(
\begin{matrix}
6149.2   & -13.6    & 0\\
-13.6    & 6169.5   & 0\\
  0       & 0        & 6093.6
\end{matrix}
\right)},\quad{\mathcal{E}_{\Omega_b}^{GBE}=\left(
\begin{matrix}
6249.0   & -2.1     & 0\\
-2.1      & 6269.4   & 0\\
  0        & 0        & 6207.6
\end{matrix}
\right)}.
\label{M_3}
\end{split}
\end{equation}}}

{\scriptsize{\begin{equation}\begin{split}
&{\mathcal{E}_{\Lambda_c}^{OGE}=\left(
\begin{matrix}
5843.4    & 45.5     & 0\\
45.5      & 5843.4   & 0\\
  0        & 0         & 5966.6
\end{matrix}
\right)},\quad{\mathcal{E}_{\Sigma_c}^{OGE}=\left(
\begin{matrix}
5855.4    & 90.9     & 0\\
90.9      & 5855.4   & 0\\
  0        & 0         & 5954.4
\end{matrix}
\right)},\\
&{\mathcal{E}_{\Xi_c}^{OGE}=\left(
\begin{matrix}
5984.9      & 28.5     & 0\\
28.5        & 5984.9   & 0\\
  0          & 0         & 6080.1
\end{matrix}
\right)},\quad{\mathcal{E}_{\Xi_c'}^{OGE}=\left(
\begin{matrix}
6010.8     & -57.1    & 0\\
-57.1      & 6010.8   & 0\\
  0         & 0         & 6096.4
\end{matrix}
\right)},\quad{\mathcal{E}_{\Omega_c}^{OGE}=\left(
\begin{matrix}
6121.5    & 40.4     & 0\\
40.4      & 6121.5   & 0\\
  0        & 0         & 6226.9
\end{matrix}
\right)}.
\label{M_4}
\end{split}
\end{equation}}}

\begin{table}
Table 6 Energies and coefficients for mixing between the configurations with $S = \frac12$ and $S = 3/2$ for bottom baryon states with $L = 1$.\label{coe1}
\begin{ruledtabular}
\begin{tabular}{l    *{13}{l}}
   &   &$|\Lambda_{b}\rangle_1$&$|\Lambda_{b}\rangle_2$&$|\Lambda_{b}\rangle_3$ &  &$|\Sigma_{b}\rangle_1$&$|\Sigma_{b}\rangle_2$& $|\Sigma_{b}\rangle_3$&  &$|\Xi_{b}\rangle_1$&$|\Xi_{b}\rangle_2$&$|\Xi_{b}\rangle_3$\\
\hline
GBE&5899.7& 0.834 &0.552&0    & 6035.7&0.834& -0.552&0    &6040.8& 0.957& 0.290&0      \\

   &5951.5& 0.552 &0.834&0    & 6087.4&-0.552&0.834&0    &6054.2& 0.290&0.957&0      \\

   &6090.7& 0     &0     &1    & 5974.6&0     &0     &1    &6079.1& 0    &0     &1      \\
\hline
   &   &$|\Xi_{b}^{'}\rangle_1$&$|\Xi_{b}^{'}\rangle_2$&$|\Xi_{b}^{'}\rangle_3$& &$|\Omega_{b}\rangle_1$&$|\Omega_{b}\rangle_2$& $|\Omega_{b}\rangle_3$&  &   & \\
\hline
   &6142.4& 0.894 &-0.447&0     &6248.8 & 0.994 &-0.103&0     &   &  &  \\

   &6176.3&-0.447 &0.894&0     &6269.6&-0.103 &0.994&0     &   &    & \\

   &6093.6&0       &0      &1    & 6207.6  &0      &0      &1       \\
\hline\hline
   &   &$|\Lambda_{b}\rangle_1$&$|\Lambda_{b}\rangle_2$&$|\Lambda_{b}\rangle_3$ &  &$|\Sigma_{b}\rangle_1$&$|\Sigma_{b}\rangle_2$& $|\Sigma_{b}\rangle_3$&  &$|\Xi_{b}\rangle_1$&$|\Xi_{b}\rangle_2$&$|\Xi_{b}\rangle_3$\\
\hline
OGE&5797.9 &0.707 &-0.707&0    & 5764.5 & 0.707 & -0.707&0     & 5956.4 & 0.707 & -0.707&0     \\

   &5888.8 &0.707 & 0.707&0    & 5946.3 & 0.707 & 0.707&0     & 6013.5 & 0.707 & 0.707&0      \\

   &5966.6& 0     &0     &1    & 5954.4&0     &0     &1    &6080.1& 0    &0     &1      \\
\hline
   &   &$|\Xi_{b}^{'}\rangle_1$&$|\Xi_{b}^{'}\rangle_2$&$|\Xi_{b}^{'}\rangle_3$& &$|\Omega_{b}\rangle_1$&$|\Omega_{b}\rangle_2$& $|\Omega_{b}\rangle_3$&  &   & \\
\hline
   &5953.7& 0.707 & -0.707&0     &6081.1& 0.707 & -0.707&0     &   &  &  \\

   &6067.9& 0.707 & 0.707&0     & 6141.3 & 0.707 & 0.707&0     &   &    &  \\

   &6096.4&0       &0      &1    & 6226.9  &0      &0      &1       \\
\end{tabular}
\end{ruledtabular}
\end{table}

Intriguingly, from Table 6, we can also see the spins of the lowest energy states in the GBE model for $\Sigma_b$, $\Xi_b$ and $\Omega_b$ are $S = 3/2$. The same phenomenon has been found for charmed baryon states. This is the most special aspect of the GBE model compared with the OGE model.

\section{Conclusions}
\label{sec:Con}

In this paper, we studied the difference between two hyperfine interactions including the Goldstone boson exchange (GBE) and the one gluon exchange (OGE) hyperfine interaction models, by predicting the masses of charmed and bottom baryons with $L = 1$ negative parity. The results for the $L = 1$ negative parity charmed and bottom baryon masses were obtained from the mass splittings between these states and their corresponding ground states, and the input parameters were determined by fitting the experimental baryon masses.

With these two models, we first expressed the fine structure correction parts ($\delta M$) for the light, charmed and bottom ground baryons as listed in Table 1 and Table 2. Then, with the mass splittings between ground baryon states, the parameters for these two kinds of hyperfine interaction models were extracted and listed in Table 3. We predicted the mass of $\Omega_b^*$, the only $L = 0$ baryon state which has not be observed, to be 6069.2 MeV. After that, the masses of the negative parity charmed and bottom baryon configurations with $L = 1$ were estimated from the splittings between their corresponding charmed and bottom baryon states with $L = 0$. In our calculations, mixing between the configurations with same spin quantum numbers was also taken into account. Then the physical masses of the negative parity charmed and bottom baryon states with $L = 1$ were predicted after diagonalizing the matrices (\ref{GBE results}), (\ref{OGE results}), (\ref{M_3}) and (\ref{M_4}) and we gave the corresponding coefficients for the mixing between these configurations in the two hyperfine interaction models.

From the latest ``Review of Particle Physics'' \cite{Agashe:2014kda}, the splitting between $\Lambda_c(2595)^+$ with $J^P = \frac12^-$ and $\Lambda_c(2625)^+$ with $J^P = \frac32^-$ is 35.86 MeV, and that between $\Xi_c(2790)^+$ with $J^P = \frac12^-$ and $\Xi_c(2815)^+$ with $J^P = \frac32^-$ is 27.5 MeV. For $\Lambda_b(5912)$ and $\Lambda_b(5920)$ with $J^P = \frac12^-$ and $J^P = \frac32^-$, respectively, $M_{\Lambda_b(5912)}$ = 5912.11 MeV, $M_{\Lambda_b(5920)}$ = 5919.81 MeV, and the mass difference between these two states is small ($<$ 8 MeV). This indicates that the spin-orbital interaction has a smaller impact on the masses of the charmed and bottom baryons with increase of the constituent quark masses. So in our calculation we neglected the contribution from the spin-orbital interaction.

It is very interesting that in the GBE model, there exist three multiplets ($\Sigma_{c(b)}$, $\Xi_{c(b)}^{'}$ and $\Omega_{c(b)}$), of which the spins of their lowest energy states are 3/2. However, in the OGE model there is no such phenomenon. There are also no such phenomenon for singly heavy baryons in QCD-motivated relativistic quark model \cite{Ebert:2011kk} and hypercentral constituent quark model \cite{Shah:2016mig,Shah:2016nxi}, and for $\Omega_c$ states in chiral quark model \cite{Wang:2017hej} and
nonrelativistic constituent quark model \cite{Chen:2017gnu}. This is the most obvious difference between the GBE and OGE models. According to our analysis, we find the contributions from the diagonal matrix elements of fine structure corrections to the energies of spin $1/2$ states are larger than the contributions to the energies of the $S = 3/2$ states only in the GBE model, as listed in matrices (\ref{GBE hyperfine}) and (\ref{OGE hyperfine}). Another obvious feature is that the mixing in the case of the bottom baryon states is stronger than that in the charmed baryon states both in the GBE and OGE models, as listed in Tables 5 and 6. This is because the larger constituent mass of the $b$ quark reduces the hyperfine interaction contributions to bottom baryon states compared with the charmed baryon states. We expect that our results for these two different models in this work can be tested at the LHC and other experiments in the near future.

The predicted masses in this paper may be also useful for the discovery of the unobserved charmed and bottom baryon states and the $J^P$ assignment of these baryon states when they are observed in the near future. It will also allow us to compare  these two different hyperfine interaction models from their results and examine which phenomenological model can better describe the spectra. Therefore, more efforts should be given to study charmed and bottom baryons both theoretically and experimentally.

\acknowledgments
This work was supported by National Natural Science Foundation of China under contract numbers 11175020, 11575023 and U1204115.


\end{CJK*}
\end{document}